\documentclass[a4paper]{jpconf}

\usepackage{graphicx}

\begin{document}

\title{Recent STAR results in high-energy polarized proton-proton collisions at RHIC}

\author{Bernd Surrow for the STAR Collaboration}

\address{Massachusetts Institute of Technology, 77 Massachusetts Avenue, Cambridge, MA 02139, USA}

\ead{surrow@mit.edu}


\begin{abstract}
The STAR experiment at the Relativistic Heavy-Ion Collider at Brookhaven National
Laboratory is carrying out a spin physics program in high-energy polarized $\vec{p}+\vec{p}$ collisions 
at $\sqrt{s}=200-500\,$GeV
to gain a deeper insight into the spin structure and
dynamics of the proton. 

One of the main objectives of the spin physics program at RHIC is the extraction of
the polarized gluon distribution function based on measurements of gluon initiated processes, such
as hadron and jet production. 
The STAR detector is well suited for the reconstruction of various final states 
involving jets, $\pi^{0}$, $\pi^{\pm}$, e$^{\pm}$ and $\gamma$, which allows to measure several different processes. 
Recent results will be shown on the measurement 
of jet production and hadron production at $\sqrt{s}=200\,$GeV.

The RHIC spin physics program has recently completed the first data taking period in 2009 of 
polarized $\vec{p}+\vec{p}$ collisions at $\sqrt{s}=500\,$GeV. This opens a new era in the 
study of the spin-flavor structure of the proton based on the production of $W^{-(+)}$ bosons.
Recent STAR results on the first measurement of $W$ boson production in polarized $\vec{p}+\vec{p}$ collisions
will be shown.
\end{abstract}


\section{Introduction}

The spin structure and dynamics of the nucleon is one of the fundamental and unresolved
questions in Quantum Chromodynamics (QCD). Contrary to atomic physics where the total
angular momentum of an atom can be accounted for by basic quantum theory in terms of its underlying atomic constituents, 
a complete theoretical description of the proton spin decomposed in terms of contributions from quark and gluon 
angular momenta and spins is still missing. Various experimental programs have been conducted
in the past to deepen our understanding on the proton spin. Deep-inelastic scattering (DIS)
experiments have clearly established that the quark spin contribution is small and
accounts for only $\approx 25\%$ of the proton spin \cite{Ashman:1989ig, Filippone:2001ux}.

High energy polarized $\vec{p}+\vec{p}$ collisions at $\sqrt{s}=200-500\,$GeV at RHIC provide a unique way to probe
the proton spin structure and dynamics using hard scattering processes \cite{Bunce:2000uv}. The production of jets and
hadrons is the prime focus of the gluon polarization studies. The production of $W^{-(+)}$ bosons at $\sqrt{s}=500\,$GeV
provides an ideal tool to study the spin-flavor structure of the proton. This has been pointed 
out at the very early design stages of the polarized proton collider program at RHIC \cite{Underwood:1991ak}.

\begin{figure}[t]
\centerline{\includegraphics[width=135mm]{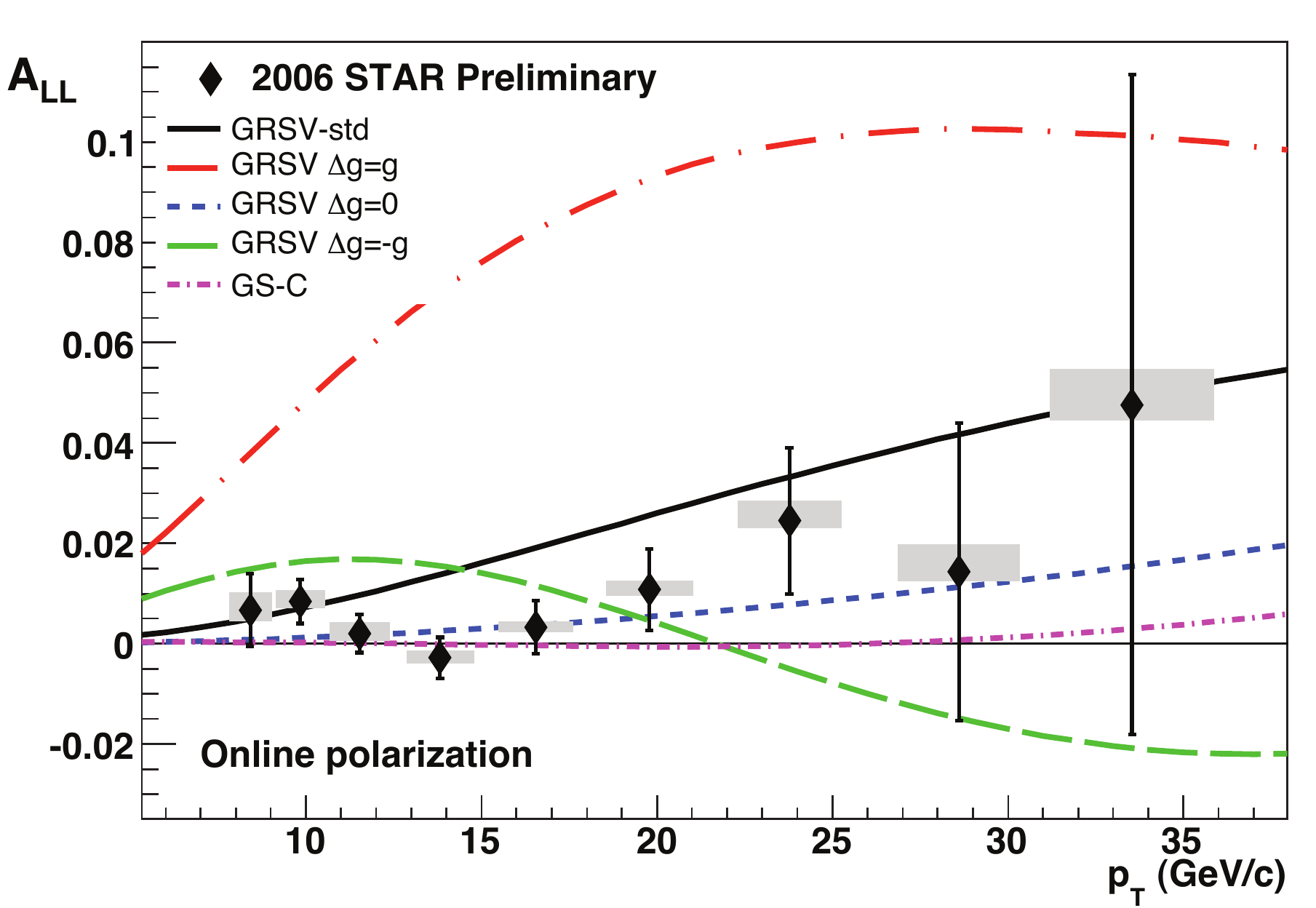}}
\label{Jets}
\caption{{\it Recent STAR 2006 $A_{LL}$ inclusive jet result.}}
\end{figure}

The first global analysis of polarized DIS data, as well as results obtained by 
the PHENIX \cite{Adare:2007dg} and STAR \cite{Abelev:2007vt} experiments in polarized $\vec{p}+\vec{p}$ collisions at RHIC placed a strong constraint on the 
gluon spin contribution in the gluon momentum range of $0.05<x<0.2$, and suggested that the gluon spin contribution is not large
in that range \cite{deFlorian:2008mr}. Constraining the polarized gluon distribution function,
$\Delta g$, through inclusive measurements has been, so far, the prime
focus of the STAR physics analysis program of the Run
3/4~\cite{Abelev:2006uq}, Run 5~\cite{Abelev:2007vt} and Run 6~\cite{ref_spin2008_murad} data samples.
Inclusive measurements, such as inclusive jet production, integrate
over a fairly large $x$ region for a given jet transverse momentum
region. While those measurements provide a strong constraint on the
value of $\Delta g$ integrated over a range in $x$, those measurements do not
permit a direct sensitivity to the actual $x$ dependence. This 
motivates the need for correlation measurements in polarized $\vec{p}+\vec{p}$ collisions.

The STAR collaboration has presented a first measurement of the longitudinal spin transfer 
$D_{LL}$ in inclusive $\Lambda$ ($\Lambda\rightarrow p \pi^{-}$) and $\bar{\Lambda}$ ($\bar{\Lambda}\rightarrow \bar{p}\pi^{+}$) 
production in polarized proton-proton collisions at a center-of-mass 
energy of $\sqrt{s}=200\,$GeV \cite{Abelev:2009xg}. This measurement may provide constraints on strange (anti) quark polarization \cite{Xu:2005ru} and can yield new insight into
polarized fragmentation functions. 

The first data taking period in 2009 of polarized $\vec{p}+\vec{p}$ collisions at $\sqrt{s}=500\,$GeV opens a new era in the 
study of the spin-flavor structure of the proton based on the production of $W^{-(+)}$ bosons. $W^{-(+)}$ bosons are
produced predominantly through $\bar{u}+d$ $(u+\bar{d})$ collisions and can be detected through
their leptonic decay. Quark and anti-quark helicity distribution functions are probed at large
scales set by the mass of the $W$ boson ($Q \sim m_{W}$) where theoretical calculations are well under control. 
The theoretical framework is well developed to describe the 
production of $W$ bosons in high-energy polarized $\vec{p}+\vec{p}$ collisions including the description of
the $W$ decay leptons subject to experimental cuts, such as the transverse momentum $p_{T}$ and pseudo-rapidity
$\eta$ of the final-state leptons \cite{Nadolsky:2003ga, deFlorian:2010aa}. This development profits from a rich history of hadroproduction of weak bosons 
at the CERN SPS and the FNAL Tevatron and provides a firm basis to use
$W$ production as a new high-energy probe in polarized $\vec{p}+\vec{p}$ collisions \cite{Kotwal:2008zz}.

The STAR Collaboration has provided critical measurements on the sudy of transverse spin
effects in polarized ~p+~p collisions at RHIC. An overview of these results is discussed in a separate
paper \cite{ref_ww2010_len}.

\section{Recent jet $A_{LL}$ results}

\begin{figure}[t]
\centerline{\includegraphics[width=115mm]{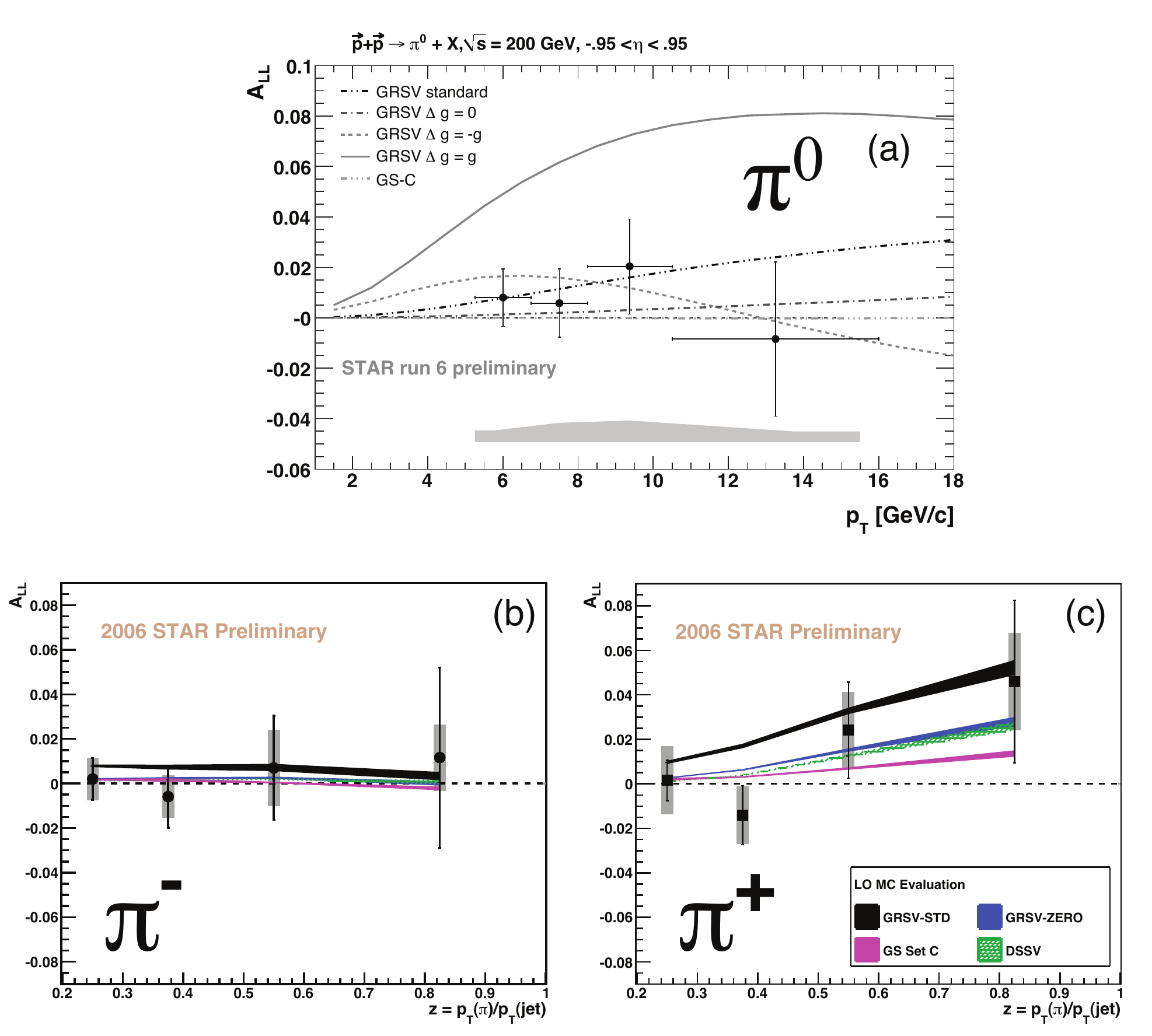}}
\label{hadrons}
\caption{{\it Recent STAR 2006 $A_{LL}$ neutral pion result (a) and STAR 2006 $A_{LL}$ results for charged pion ($\pi^{-}$ and $\pi^{+}$) / jet correlation measurements (b/c).}}
\end{figure}

STAR reconstructs jets with the midpoint cone algorithm using clusters
of charged track momenta measured with the STAR Time Projection Chamber (TPC) and
tower energy deposits in the STAR Barrel Electromagnetic Calorimeter
(BEMC) within a cone radius of $R\equiv\sqrt{\Delta
\eta^{2}+\Delta\phi^{2}}\;$\cite{ref_spin2008_murad}. The jet sample 
was required to be within a fiducial range of
$-0.7<\eta_{\rm{Detector}}<0.9$ for the 2006 data sample
with a cone radius of $R=0.7$. The dominant fraction of jet
events in the 2006 data sample are based on a jet patch
(JP) trigger that required a minimum energy deposition for a group of
towers over a region of $\Delta \eta \times
\Delta \phi = 1.0 \times 1.0$.  This trigger was taken in coincidence
with a minimum-bias condition using the STAR Beam-Beam Counter
(BBC). 

\begin{figure}[t]
\centerline{\includegraphics[width=130mm]{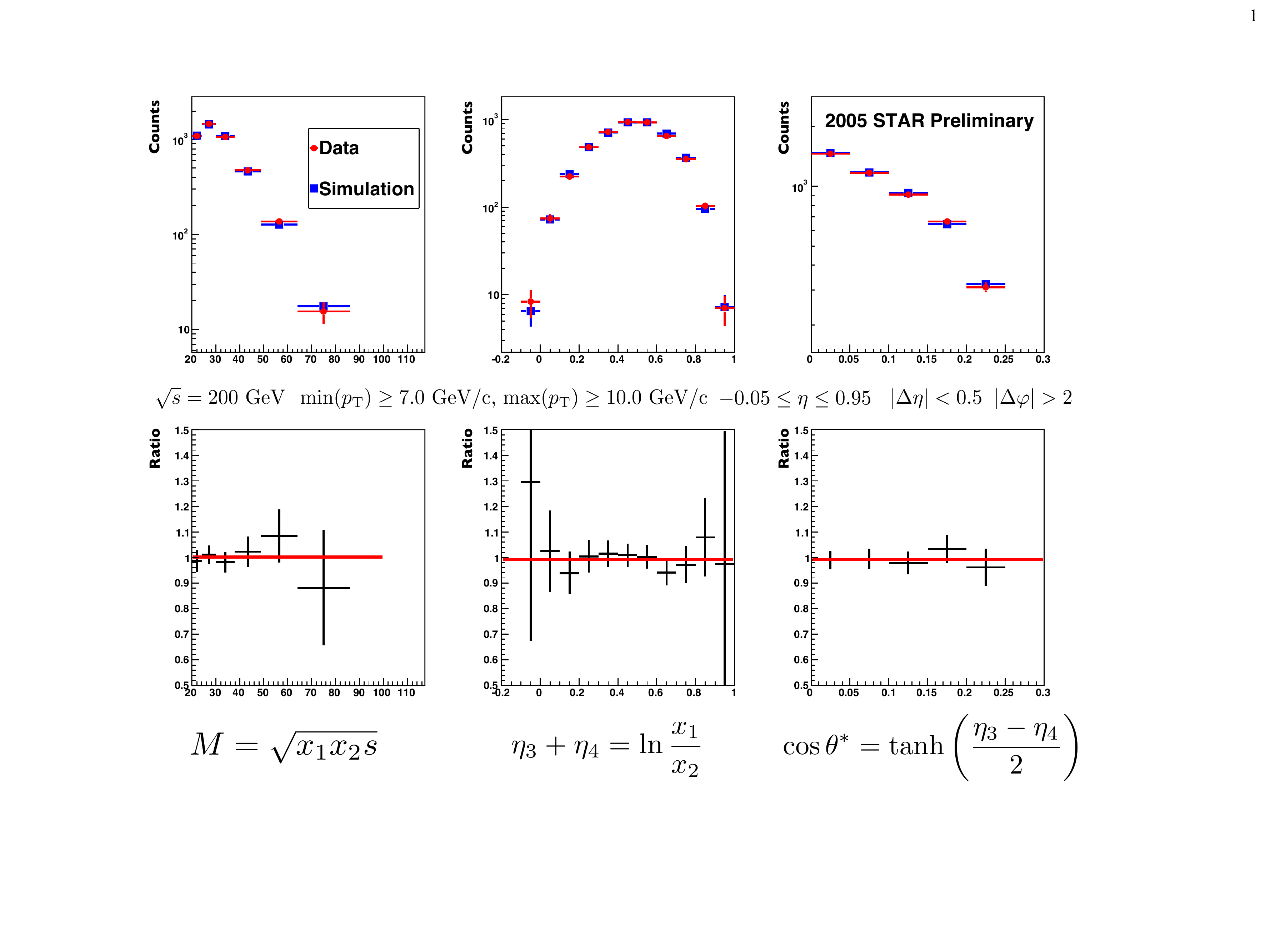}}
\label{data_mc-di-jets}
\caption{{\it Comparison of STAR 2005 di-jet data and a PYTHIA Monte-Carlo showing the yield at the top and the ratio at the bottom for three
characteristic di-jet variables (See text for further details).}}
\end{figure}

Throughout the following discussion, four gluon polarization scenarios have been used as input to NLO perturbative QCD
calculations of $A_{LL}$.
The GRSV standard case refers to a global analysis fit of polarized DIS data \cite{Gluck:2000dy}. The case 
for a vanishing gluon polarization (GRSV-ZERO) and the case of a maximally positive (GRSV-MAX) or negative (GRSV-MIN) 
gluon polarization have also been considered. 
The GS-C \cite{Gehrmann:1995ag} set of $\Delta g$ is reflected by a large positive gluon
polarization at low $x$, a node around $x\sim 0.1$ and a negative
gluon polarization at large $x$ at a scale of $Q^{2} \sim 1\,$GeV$^{2}$.

Figure \ref{Jets} shows the most recent STAR preliminary result of $A_{LL}$ for inclusive jet production as a function
of $p_{T}$ based on the 2006 data sample of $4.7\,$pb$^{-1}$.
Typical beam polarization values during the 2006 data-taking period were $55-60\%$.
The leading systematic uncertainty comes from the trigger and jet reconstruction biases. 
Differences between observed and true jet $p_T$ are estimated using PYTHIA and GEANT simulations and 
result in corrections being applied to the jet $p_T$ values.
Additionally, the trigger conditions at STAR can artificially bias the analyzed data sample towards
particular flavors of partonic collisions.  
A conservative systematic uncertainty is evaluated to account for this effect which incorporates all allowable 
gluon polarization scenarios.  
The $A_{LL}$ curves in Figure \ref{Jets} are derived from NLO fits to world polarized DIS data 
by two separate theory groups, GRSV \cite{Gluck:2000dy} and GS \cite{Gehrmann:1995ag}. The currently measured result
is consistent with a small gluon polarization scenario for the STAR x-range of $0.03<x<0.3$. This x-range 
accounts for $\sim$ 50\% of the total $\Delta G$ integral, i.e. $\Delta G=\int_{0}^{1}\Delta g dx$, for GRSV-STD.

\begin{figure}[t]
\centerline{\includegraphics[width=150mm]{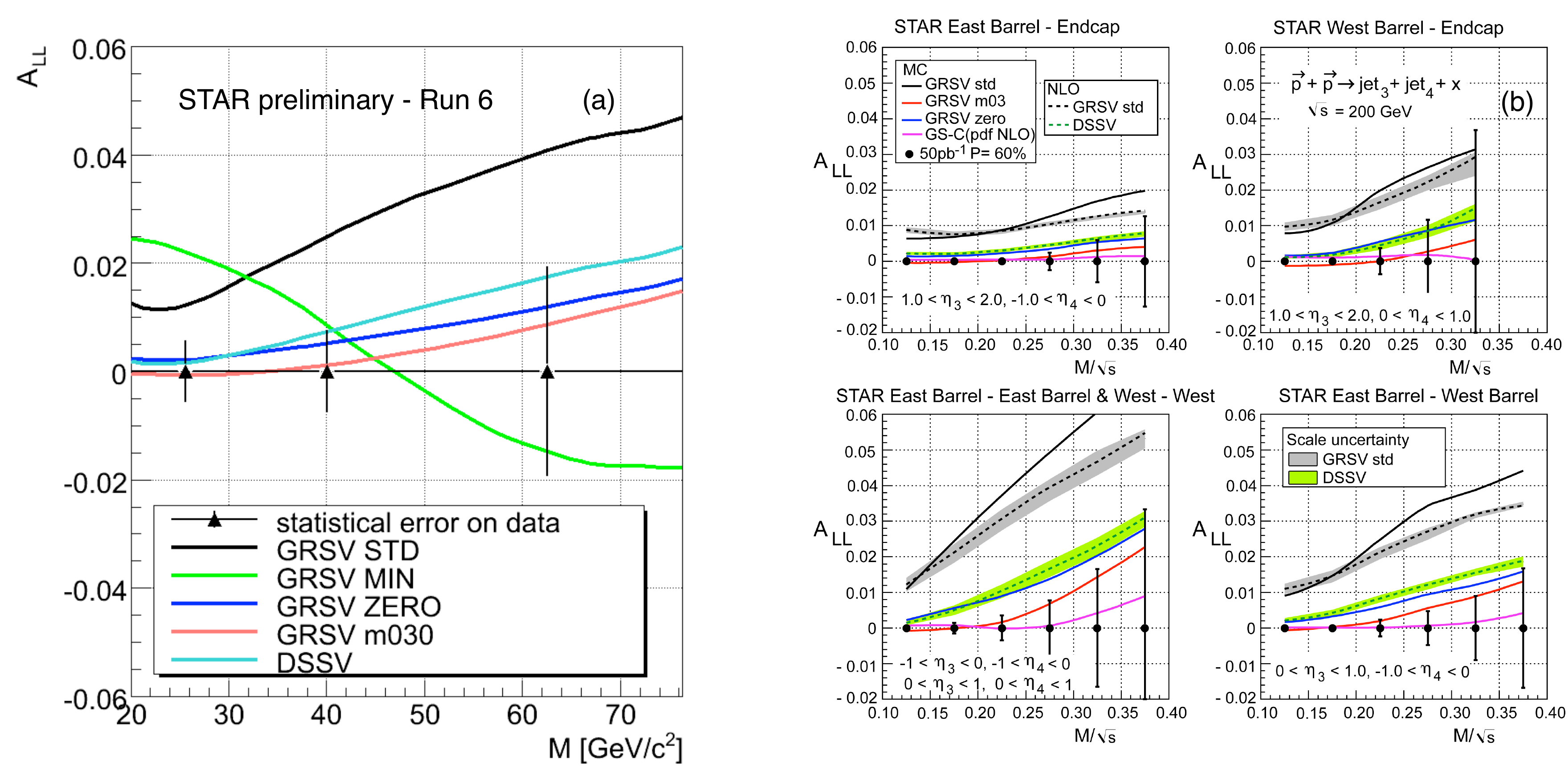}}
\label{dijets}
\caption{{\it (a) Statistical precision of the longitudinal double-spin 
asymmetry, $A_{LL}$, as a function of the di-jet invariant mass, $M$, for the 2006 RHIC data sample. (b) Longitudinal 
double-spin asymmetry, $A_{LL}$, for di-jet production as function of the ratio $M/\sqrt{s}$
for different topological combinations of the STAR BEMC and the STAR EEMC acceptance region.}} 
\end{figure}

\section{Recent neutral and charged pion $A_{LL}$ results}

Neutral pions are identified via the two-photon decay channel, which accounts
for more than $98\%$ of $\pi^{0}$ decays. Two primary sub-detectors, the STAR Barrel Electromagnetic Calorimeter (BEMC) and respective Shower
Maximum Detector (SMD) are employed to reconstruct decay photons from mid rapidity neutral pions.
Charged
pion production is of particular interest since the difference of the longitudinal double-spin asymmetries for $\pi^{+}$ and $\pi^{-}$ 
production, $A_{LL}(\pi^{+})$ - $A_{LL}(\pi^{-})$,  
tracks the sign of $\Delta g$, due to the opposite signs of 
the polarized distribution functions for up and down quarks.  
The STAR Time Projection Chamber (TPC)  offers robust reconstruction and identification of 
charged pions up to a transverse momentum of $15$ GeV/c.  Particle identification is accomplished 
using measurements of ionization energy loss in the TPC.  

Figure~\ref{hadrons} (a) shows $A_{LL}$ for neutral pion production at mid rapidity as a function of $p_{T}$ based on the 2006 data 
sample~\cite{ref_spin2008_alan}. This new measurement is shown 
in comparison to NLO calculations assuming different 
gluon polarization scenarios~\cite{ref_spin2008_murad, ref_spin2008_alan}. This measurement reaches to higher $p_{T}$ values compared
to previous measurements by PHENIX~\cite{Adare:2008px} and STAR~\cite{Abelev:2009pb}. At the present level of precision, the 
data excludes extreme gluon polarization scenarios. The STAR
collaboration has recently presented also results for neutral pion production at forward rapidity based on the STAR Electromagnetic 
Endcap Calorimeter (EEMC) and
the STAR Forward Pion Detector (FPD) \cite{ref_spin2008_scott}. These measurements provide an important milestone for future photon-jet coincidence 
measurements at STAR.

Figure~\ref{hadrons} (b) and (c) show STAR's new preliminary result of $A_{LL}$ for 
charged pion ($\pi^{-}$ and $\pi^{+})\,/\,{\rm jet}$ correlation measurements at mid rapidity. Charged pions are 
reconstructed opposite a jet that triggered the experiment. The data are shown versus
$z\equiv p_{T}(\pi)/p_{T}(jet)$ and were obtained in 2006~\cite{ref_spin2008_adam}. Full 
NLO calculations for this observable have recently been released \cite{deFlorian:2009fw}. These
calculations will provide the basis for charged pion results to be included in a global analysis. The data
shown for charged pion production are compared to a LO MC evaluation of $A_{LL}$ excluding extreme scenarios of $\Delta g$. 
The measurement of $A_{LL}(\pi^{+})$ is of particular interest, since its analyzing power is large because of the large $ug$ scattering 
contribution to 
the production cross section and sizable $\Delta u/u$.
In the future, more precise measurements of $A_{LL}(\pi^{+})$ have a great potential for providing a better understanding of $\Delta g$.

\section{Status and prospects of di-jet production}

Correlation measurements such as those for di-jet production allow for
a better constraint of the partonic kinematics and thus the shape of
$\Delta g$. At LO, the di-jet invariant mass, $M$, is proportional to
the product of the $x$ values of the partons,
$M=\sqrt{s}\sqrt{x_{1}x_{2}}$, whereas the pseudo-rapidity sum of the
final-state jets, $\eta_{3}+\eta_{4}$, is proportional to the
logarithm of the ratio of the $x$ values,
$\eta_{3}+\eta_{4}=\ln\left(x_{1}/x_{2}\right)$.  Photon-jet
coincidence measurements are expected to provide a theoretically clean
way to extract $\Delta g$. A LO extraction of $\Delta g$ alone would
allow a model-independent way to constrain the $x$ dependence, which
would be an important contribution, without making an a priori
assumption on the functional form of $\Delta g$ as is currently
required in a global analysis. 
Measurements at both $\sqrt{s}=200\,$GeV and
$\sqrt{s}=500\,$GeV are preferred to maximize the kinematic reach in
$x$ and possibly provide a means to observe effects of scaling
violations at fixed $x$ by measuring different $p_{T}$ values. The
wide acceptance of the STAR experiment permits reconstruction of
di-jet events with different topological configurations,
i.e. different $\eta_{3}$/$\eta_{4}$ combinations, ranging from
symmetric ($x_{1}=x_{2}$) partonic collisions to asymmetric
($x_{1}<x_{2}$ or $x_{1}>x_{2}$) partonic collisions. This, together
with the variation of the center-of-mass energy, is expected to constrain $\Delta g$
over a wide range in $x$ of approximately $\sim 2\cdot 10^{-3} < x <
0.3$ for di-jet and photon-jet events.  The NLO framework for correlation measurements exists and
therefore those measurements can be used in a global analysis~\cite{deFlorian:1998qp}.

Figure 3 shows a comparison of STAR 2005 di-jet data and a PYTHIA Monte-Carlo simulation with the yield at the top and the ratio at the bottom for 
the invariant mass $M$, the pseudo-rapidity sum $\eta_{3}+\eta_{4}$ and the cosine of the partonic center-of-mass scattering angle $\cos\theta^{*}$ determined from
the pseudo-rapidities of both final-state jets. The normalization between data and Monte-Carlo is fixed by the invariant mass distributions. The same normalization
factor is then applied to both other di-jet distributions. Data and Monte-Carlo are found to be in good agreement.

\begin{figure}[t]
\centerline{\includegraphics[width=160mm]{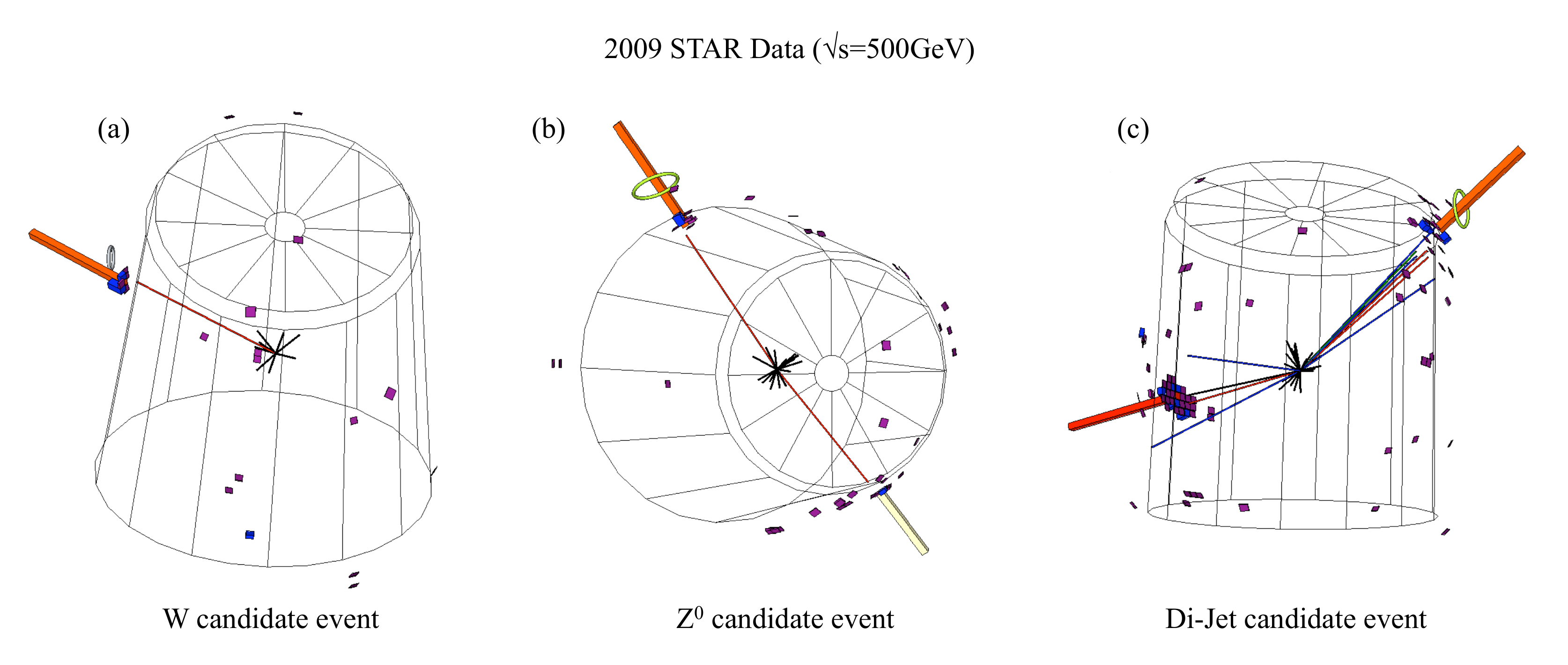}}
\label{event-display-500}
\caption{{\it Event displays from the 2009 data taking period at $\sqrt{s}=500\,$GeV showing a $W$ candidate event (a), a $Z^{0}$ candidate event and a di-jet candidate event (c).}}
\end{figure}

\begin{figure}[t]
\centerline{\includegraphics[width=120mm]{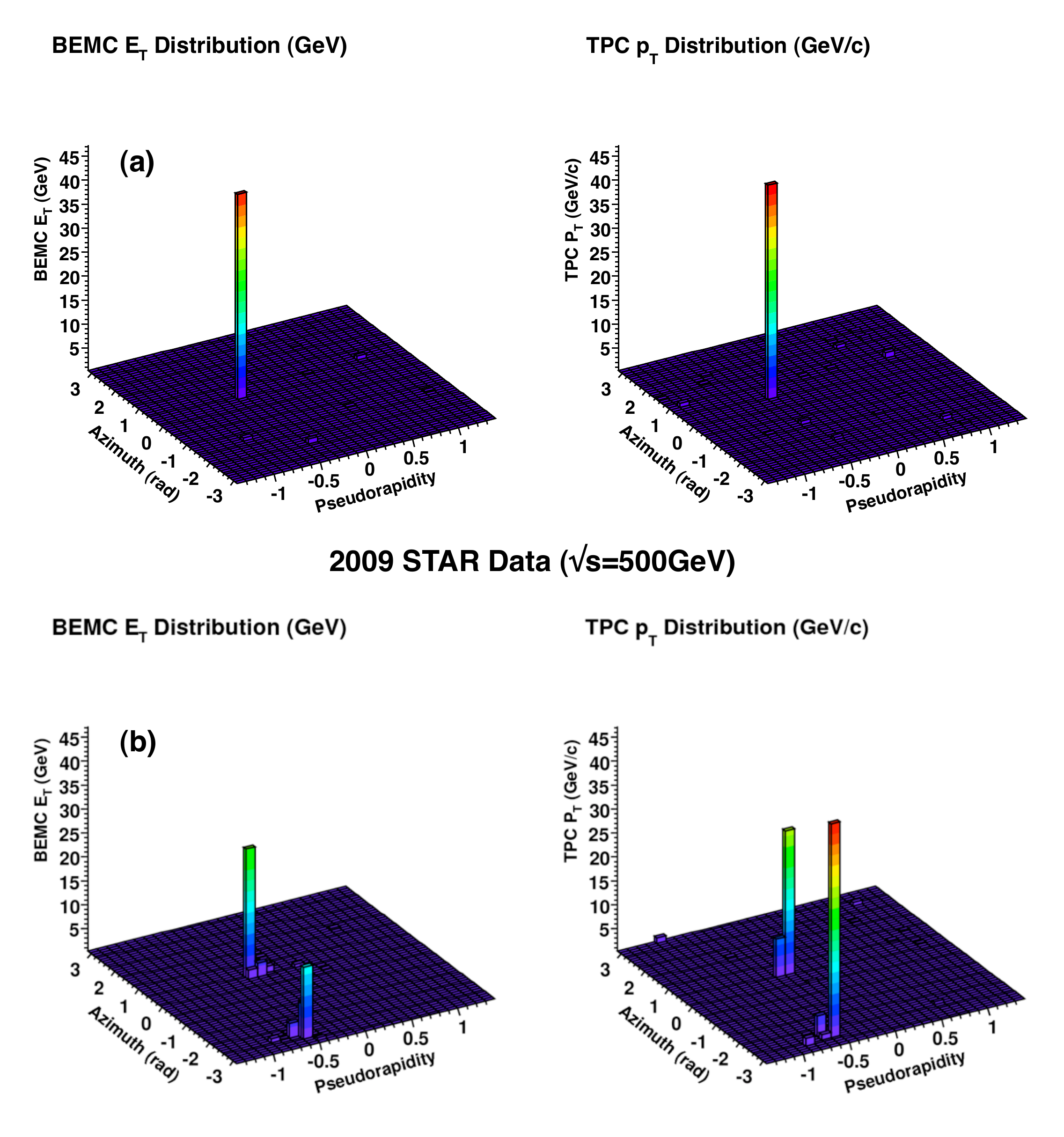}}
\label{lego-500}
\caption{{\it Lego plot showing the BEMC $E_{T}$ and the TPC $p_{T}$ distribution for a $W$ candidate event (a)
and a di-jet candidate event (b).}}
\end{figure}

Figure 4 (a) shows the statistical precision of the
longitudinal double-spin asymmetry, $A_{LL}$, as a function of the
di-jet invariant mass, $M$. Those uncertainties, extracted from the
2006 data sample, are compared to a LO MC evaluation of $A_{LL}$
computed with a PYTHIA MC sample using different event weights to
account for different polarized gluon distribution functions of
GRSV~\cite{Gluck:2000dy} and DSSV~\cite{deFlorian:2008mr}. The size of the
statistical uncertainty at the highest invariant mass bin is at the
level of the difference between GRSV-STD and DSSV.

\begin{figure}[t]
\centerline{\includegraphics[width=150mm]{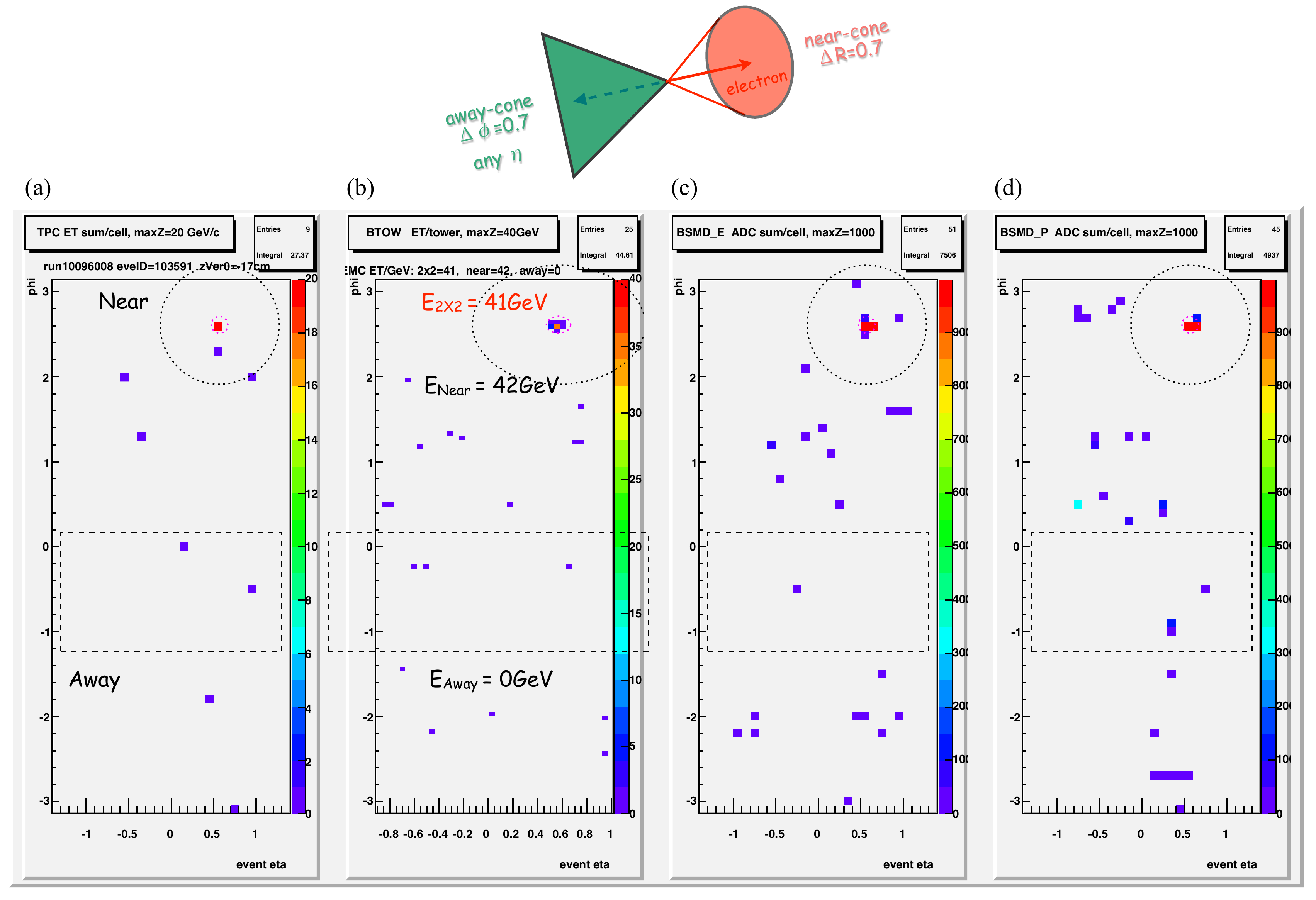}}
\label{algo}
\caption{{\it Illustration of the $W$ offline selection criteria (top) and the distribution of the TPC $p_{T}$ (a) the BEMC $E_{T}$ (b) and both BEMC SMD layers (c) / (d) in azimuth 
and pseudo-rapidity.}}
\end{figure}

Figure 4 (b) shows the expected precision for the longitudinal double-spin
asymmetry, $A_{LL}$, for di-jet production as a function of
$M/\sqrt{s}$ for different topological combinations of the STAR BEMC
and the STAR Endcap Electromagnetic Calorimeter (EEMC) acceptance
regions. Taking into account the different $\eta$ ranges
covered, and equivalently, the different $\cos\theta^{*}$ regions
being probed, each panel represents a different range in
$x_{1}$/$x_{2}$. At LO, $\cos\theta^{*}$ amounts to
$\tanh\left(\frac{\eta_{3}-\eta_{4}}{2}\right)$. The upper left panel
effectively probes asymmetric partonic collisions where predominantly
a low-x gluon collides with a high-x quark at large invariant
masses. 
\begin{figure}[t]
\centerline{\includegraphics[width=110mm]{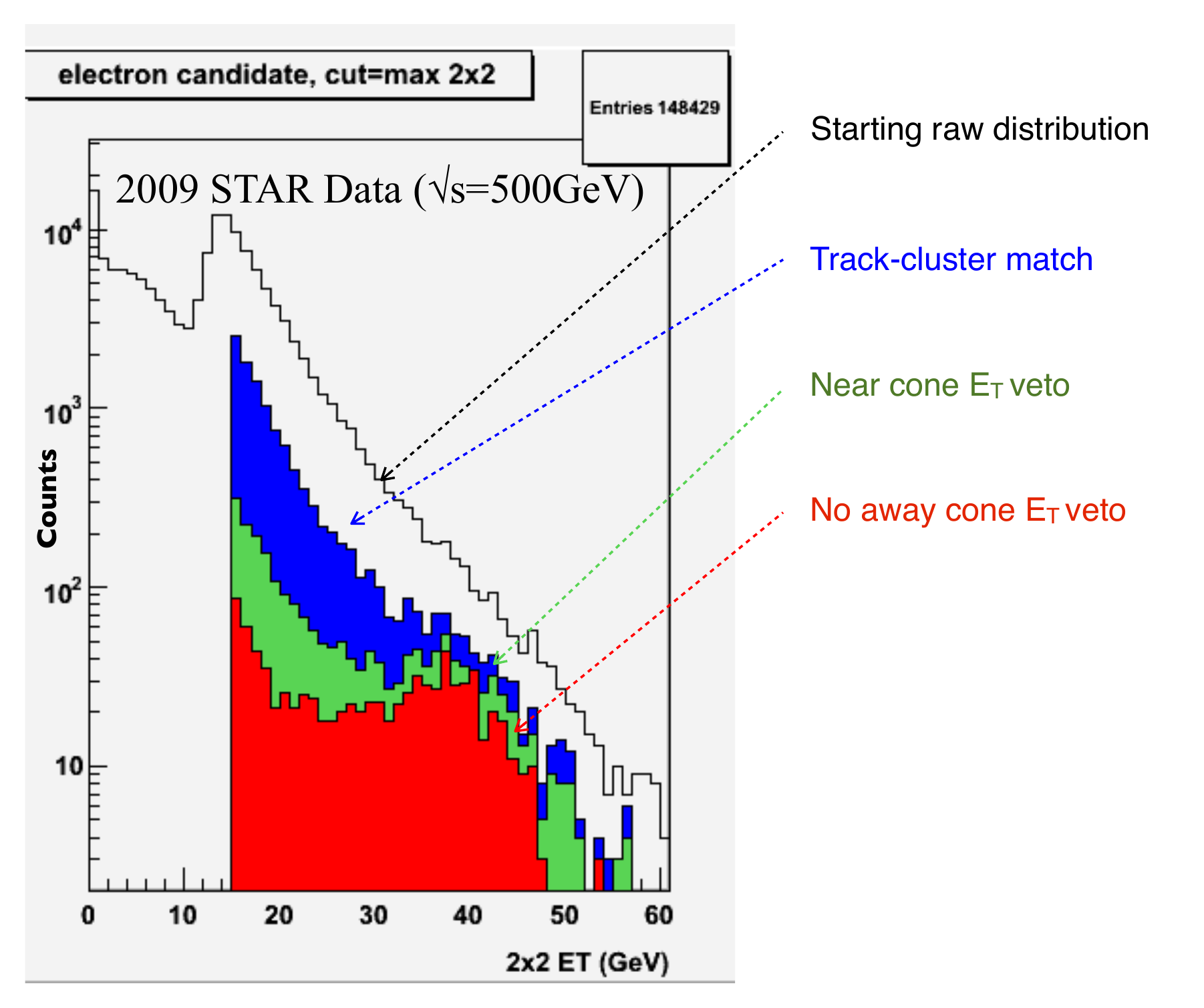}}
\label{w-cuts}
\caption{{\it Evolution of $W$ offline selection cuts (See text for further details).}}
\end{figure}
The effective variation
of $x_{1}$ and $x_{2}$ amounts to $0.2<x_{1}<0.6$ and
$0.07<x_{2}<0.2$. In contrast, a kinematic region of larger $x$ values
is probed at predominantly symmetric partonic collisions such as the one
shown in the lower right panel.
The effective variation of $x_{1}$ and $x_{2}$ is roughly
equal and given by the horizontal axis of the lower right panel.  The
projected uncertainties are shown for a luminosity of $50\,$pb$^{-1}$
and a beam polarization of $60\%$. Those projected uncertainties are
compared to a LO evaluation of $A_{LL}$ and a full NLO $A_{LL}$
calculation. Scale uncertainties are shown as a shaded band for DSSV
and GRSV-STD reflecting a variation of the invariant mass $M$ as a
hard scale of $2M$ and $0.5M$. Asymmetric cuts are imposed for the LO
MC and the NLO determination of $\rm{min}(p_{T})\geq 7\,$GeV/c and
$\rm{max}(p_{T})\geq 10\,$GeV/c.  The result of a LO MC evaluation using
GS-C \cite{Gehrmann:1995ag} for $\Delta g$ is also shown. GS-C is still consistent with the
current inclusive jet results~\cite{ref_spin2008_murad}.  A cone
radius of $R=0.7$ has been used. Good agreement is found between a LO
MC evaluation of $A_{LL}$ and a full NLO calculation.  Scale
uncertainties are found to be small in comparison to the variation of
the chosen polarized gluon distribution functions, in particular, at
large values of $M$. 

While inclusive measurements from STAR have provided important constraints
on $\Delta g$, and will continue to do so, the impact of correlation
measurements to the $x$ dependence should greatly enhance our understanding
of $\Delta g$.

\section{Measurement of $W$ boson production at STAR}

The data used for the first $W$ boson production analysis at STAR were collected in 2009 colliding polarized
proton beams at $250\,$GeV. The STAR detector systems used in this analysis are the STAR Time Projection Chamber (TPC), the STAR Barrel Electromagnetic Calorimeter (BEMC) 
and STAR Electromagnetic Endcap Calorimeter (EEMC). Only their tower response has been taken into
account in this analysis. 
The BEMC was used to measure the transverse energy, $E_{T}$, of $e^\pm$. 
The suppression of the QCD background by several orders of magnitude was based on the TPC, BEMC and EEMC.
Figure 5 shows event displays from the 2009 data taking period for a $W$ candidate event (a), a $Z^{0}$ candidate event and a 
di-jet candidate event (c). A lego plot is shown in Figure 6 displaying the BEMC $E_{T}$ and the TPC $p_{T}$ distribution for a $W$ candidate event at the top 
and a di-jet candidate event at the bottom.  

Proton-proton collision events focusing on $W$ production at $\sqrt{s}=500\,$ were identified by  
a two-step energy requirement in the BEMC. Electrons and positrons from $W$ production at mid rapidity are characterized by
large $E_{T}$ peaked at $\sim M_{W}/2$ (Jacobian peak). At the hardware trigger level (L0), a high tower (HT) calorimetric trigger
condition required $E_{T}>7.3\,$GeV in a single BEMC tower. At the software trigger level (L2), a dedicated
trigger algorithm was developed that required that a $2 \times 2$ tower cluster $E_{T}$ sum exceeds $13\,$GeV.
The offline selection of W candidate events is based on kinematical and topological difference between leptonic $W^{\pm}$ decays,
and QCD background events as shown in Figure 5 and Figure 6. Events from $W^\pm\rightarrow e^\pm+\nu$ decays contain a nearly isolated 
$e^{\pm}$ with a neutrino in the opposite direction in azimuth. The neutrino escapes detection, leading 
to a large missing energy. 

Figure 7 illustrates the basic idea of the selection of $W$ candidate events. The 2D histograms show for one $W$ candidate event
the TPC $p_{T}$ (a) and the BEMC tower $E_{T}$ (b) distribution along with both BEMC shower-maximum detector (SMD) layers (c) / (d) as a function of the 
azimuthal angle $\phi$ and the pseudo-rapidity $\eta$. The approximate area of a $2 \times 2$ tower cluster 
is indicated by a small circle around the 
large energy deposition in a single BEMC tower. Also shown is the size of a cone (near-cone) around the electron candidate. The away-side area is sketched 
by the black dashed rectangular region. The TPC high-$p_{T}$ entry appears in the same $\eta-\phi$ region as the large BEMC $E_{T}$ entry. 
Both BEMC SMD layers show a large energy deposition at the same location in the $\eta-\phi$ plane as the TPC high-$p_{T}$ and the BEMC
high-$E_{T}$ entry.
An electron candidate is defined to be any TPC track with $p_T>10\,$GeV$/c$ that is associated with a primary vertex 
with $|z|<100\,$cm, where $z$ is the distance along the beam direction.
A $2\times 2$ BEMC tower cluster $E_{T}$ sum
is required to be larger than $15\,$GeV. The excess BEMC $E_{T}$ sum in a $4 \times 4$ tower cluster centered around the respective $2\times 2$ tower 
cluster is required to be below $5\%$. In addition, the distance between the $2\times 2$ cluster tower centroid and the TPC track is required to be less 
than $7\,$cm. A near-cone
is formed around the electron candidate direction with a radius in $\eta$-$\phi$ space of $R=0.7$. The excess BEMC and TPC $E_{T}$ sum is required to be less than
$12\%$ of the $2\times 2$ cluster $E_T$. The away side requirement is based on a cut on the BEMC and TPC $E_{T}$ sum to be less than $8\,$GeV. This sum is
extended over the full $\eta$ range of the BEMC and EEMC and $\Delta \phi=0.7$ as shown in Figure 7.
Figure 8 shows the evolution of all cuts for the BEMC $2 \times 2$ $E_{T}$ distribution. A clear Jacobian peak emerges characteristic for $W$
production in contrast to QCD background dominating the low $E_{T}$ region.

A data-driven
procedure was applied to estimate the level of QCD background events. 
The first part consists of an estimate of the impact of the missing calorimeter coverage for $-2<\eta<-1$ on the cluster $E_{T}$ distribution. 
This estimate was obtained by performing
a parallel analysis without the EEMC as an active detector. The difference in yield of accepted events as a function of $E_{T}$ 
provides an absolute estimate for the missing calorimetric coverage. 
The remaining 
background is estimated by fitting a separate QCD shape distribution in $E_{T}$ to the region of $E_{T}<19\,$GeV shown as the black histogram in 
Figure 9. 
The fitted QCD background distribution is shown by the
blue histogram in Figure 9. The background subtracted distribution is shown by the yellow histogram. Figure 9 also
shows the distribution from a PYTHIA Monte-Carlo simulation. Good agreement in yield and shape is found between the data and Monte-Carlo distribution for the
charge-sum BEMC $E_{T}$ distribution. In the meantime, a considerable effort has been performed on the TPC calibration to allow a full charge-sign separation. This
and in particular the first measurement of the respective charge-separated cross-sections and parity-violating singe-spin asymmetries $A_{L}$ at mid rapidity will 
be discussed in a forthcoming publication. 

Future planned high-statistics measurements at forward and 
backward pseudo-rapidities will focus on constraining the polarization of $\bar{d}$ and $\bar{u}$  quarks, respectively 
to deepen our understanding of the spin structure of the QCD sea.

\begin{figure}[t]
\centerline{\includegraphics[width=160mm]{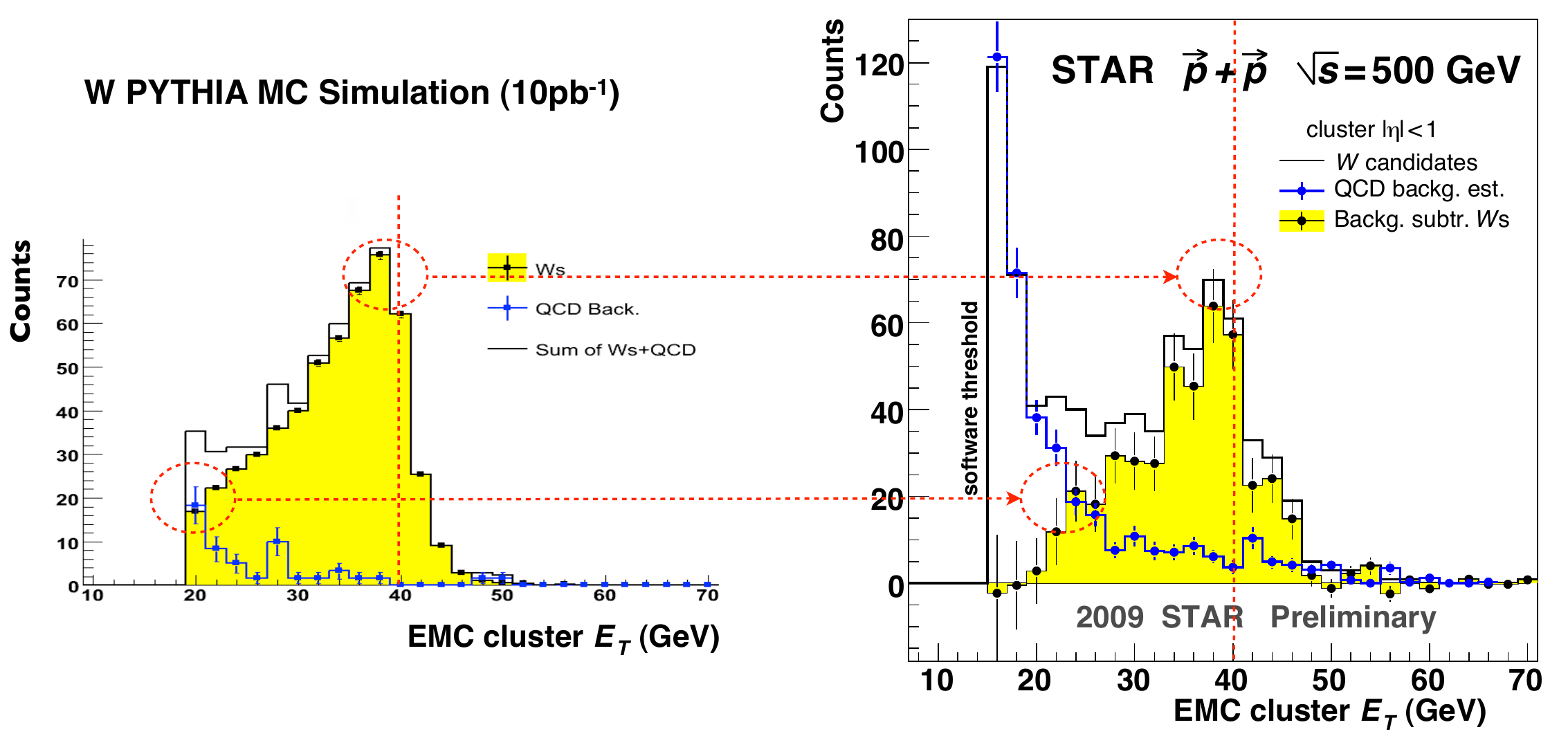}}
\label{w-jacobian}
\caption{{\it Reconstructed BEMC lepton transverse energy for $W$ events.}}
\end{figure}

\section*{References}

\bibliographystyle{iopart-num}
\bibliography{surrow}

\end{document}